# MIRAC-5: A ground-based mid-IR instrument with the potential to detect ammonia in gas giants

R. Bowens[a,*], J. Leisenring[b], M. R. Meyer[a], M. Montoya[b], W. Hoffmann[b],
K. Morzinski[b], P. Hinz[c], J. D. Monnier[a], E. Bergin[a], E. Viges[d], P. Calissendorff[a],
W. Forrest[e], C. McMurtry[e], J. Pipher[e,†], M. Cabrera[f]

[a]Department of Astronomy, University of Michigan, 1085 S. University, Ann Arbor, MI 48103;
[b]Steward Observatory, The University of Arizona, 955 N. Cherry Ave, Tucson, AZ 85719;
[c]Astronomy & Astrophysics Department, UC Santa Cruz, 1156 High St., Santa Cruz, CA 95064;
[d]Space Physics Research Lab, University of Michigan, 2455 Hayward St., Ann Arbor, MI 48109;
[e]Department of Physics and Astronomy, University of Rochester, 206 Baucsh and Lomb Hall, Rochester, NY 14627; [f]Conceptual Analytics LLC, 8209 Woburn Abbey Rd., Glenn Dale, MD 20769

## ABSTRACT

We present the fifth incarnation of the Mid-Infrared Array Camera (MIRAC-5) instrument which will use a new GeoSnap (3 – 13 microns) detector. Advances in adaptive optics (AO) systems and detectors are enabling ground-based mid-infrared systems capable of high spatial resolution and deep contrast. As one of the only 3 – 13 micron cameras used in tandem with AO, MIRAC-5 will be complementary to the James Webb Space Telescope (JWST) and capable of characterizing gas giant exoplanets and imaging forming protoplanets (helping to characterize their circumplanetary disks). We describe key features of the MIRAC-5 GeoSnap detector, a long-wave Mercury-Cadmium-Telluride (MCT) array produced by Teledyne Imaging Sensors (TIS), including its high quantum efficiency (> 65%), large well-depth, and low noise. We summarize MIRAC-5's important capabilities, including prospects for obtaining the first continuum mid-infrared measurements for several gas giants and the first 10.2-10.8 micron $NH_3$ detection in the atmosphere of the warm companion GJ 504b (Teff ~ 550 K) within 8 hours of observing time. Finally, we describe plans for future upgrades to MIRAC-5 such as adding a coronagraph. MIRAC-5 will be commissioned on the MMT utilizing the new MAPS AO system in late 2022 with plans to move to Magellan with the MagAO system in the future.

**Keywords:** Mid-Infrared, Adaptive Optics, High Contrast

## 1. INTRODUCTION

Exoplanet studies have evolved greatly over the past two decades, but more data are essential to answer pressing questions. Key to our efforts is direct imaging which allows estimates of luminosity, temperature, and atmospheric compositions of targets, enabling better constraints on formation models and subsequent planet evolution. Mid-infrared (3 – 13 microns) direct imaging is useful for covering a wide temperature range, probing through dust, and requiring less stringent contrasts compared to planet studies with reflected light. While JWST will revolutionize the field of infrared (IR) astronomy, ground-based systems supported by AO can provide greater contrast, making them complementary to JWST.

MIRAC is an evolving sequence of mid-IR cameras which began in 1988 as a collaboration between University of Arizona (UA), Smithsonian Astrophysical Observatory, and the Naval Research Laboratory[1]. MIRAC-3 operated successfully from 2000 to 2008 on MMT and Magellan. It was the first time MIRAC was mated to the Bracewell Infrared Nulling Cryostat (BLINC)[2]. MIRAC-4 which operated from 2008 to 2011 featured several advancements including a larger cryostat with mechanical pulse-tube cooling. However, MIRAC-4's 256x256 Si:As BIB array suffered from image artifacts and had limited astronomical application. MIRAC-5, a joint effort between UA and University of Michigan (UM), features a new 1024x1024 MCT array from TIS. MIRAC-5 is intended for use with MAPS on MMT[3] and eventually MagAO at Magellan[4]. With the new detector, MIRAC-5 should be capable of obtaining the first mid-IR continuum detection for several known wide-orbit gas giants. Furthermore, MIRAC-5 should be able to detect ammonia in the atmosphere of the ~550 K companion GJ504b[5]. With future improvements, including a coronagraph, MIRAC-5 will be able to image forming protoplanets and gas giants inferred via radial velocity surveys.

---

[*]rpbowens@umich.edu; https://sites.lsa.umich.edu/feps/

[†] Deceased.

In this paper, we first describe the detector array for MIRAC-5 in Section 2 before covering the cryostat and optics in Section 3. In Section 4, we discuss the design and expectations for the ammonia filter. We close with future plans in Section 5.

## 2. DETECTOR ARRAY

The baseline GeoSnap, a 2048x2048 format device, features a longwave MCT photosensitive wafer with an anti-reflection coating providing > 90% quantum efficiency (QE) at 10 microns[6,7] (c.f. McMurtry et al.[8]). As an initial test device, the MIRAC-5 GeoSnap (Figure 1) utilizes only a single 1024x1024 quadrant without anti-reflection coating, limiting its QE to 65% in the 3 to 13 micron range (Figure 2). While GeoSnap's readout integrated circuit (ROIC) is capable of full frame readout rates up to 120 Hz, MIRAC's software implementation currently limits it to 85 Hz frame rate. The MIRAC-5 GeoSnap features a well depth of 1.2 million electrons in low gain mode, whereas more recent versions of the device have approximately double the well depth. The dark current of GeoSnap is approximately 6600 e-/sec in a 38 K environment and the read noise is approximately 140 e-/pixel RMS for a single frame in low gain mode. For an 85 Hz framerate at half well, the shot noise is approximately 775 e-/pixel; dominating over the detector noise (dark plus read noise).

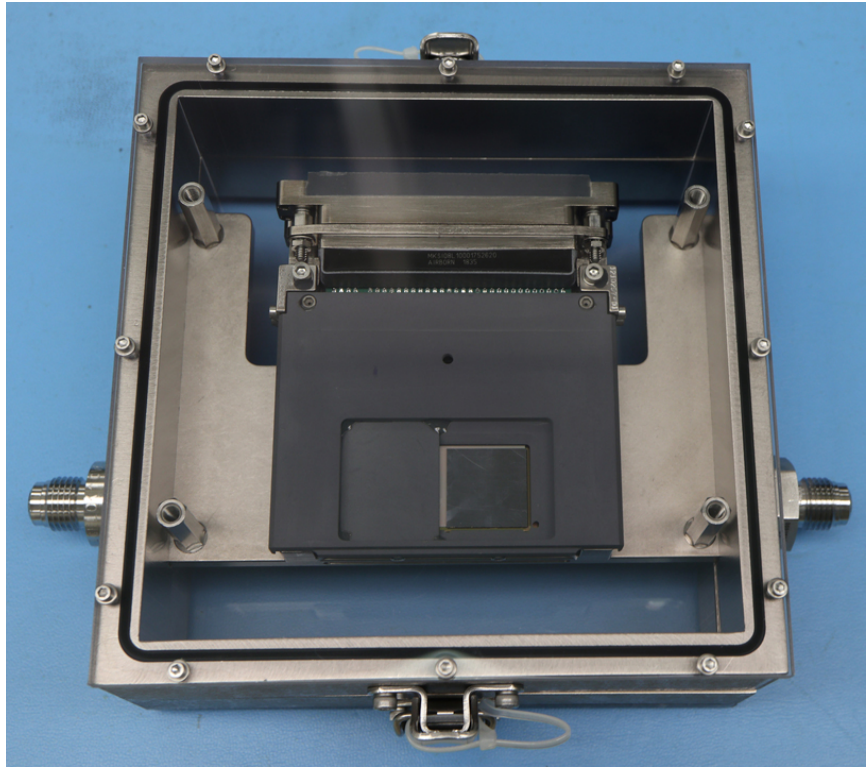

Figure 1: Close-up image of the MIRAC-5 GeoSnap detector. Note that only one of the four ROIC quadrants is exposed, which corresponds to the 1024x1024 hybridized region.

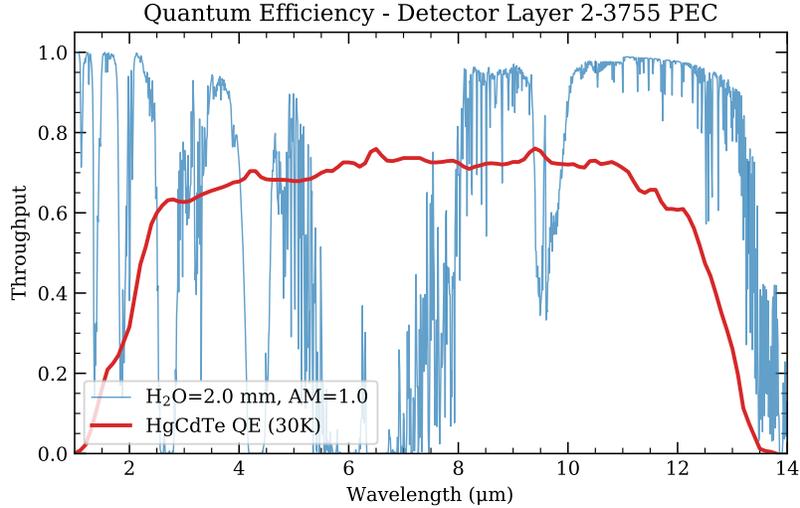

Figure 2: The quantum efficiency of the MIRAC-5 GeoSnap at 30 K in red. Atmospheric transmission for a column density of 2.0 mm precipitable water vapor at zenith (one airmass) is shown in blue[9]. The QE is provided by TIS using a process evaluation chip measurement.

The MIRAC-5 GeoSnap is unaffected by Excess Low Frequency Noise (ELFN), due to fluctuations in the space charge which can occur from the generation and recombination in the blocking layer in the Aquarius and similar arrays[10]. The fluctuating potential across the IR active layer results in randomly modulated photo-response. However, the MIRAC-5 GeoSnap does suffer from 1/f noise. The origin of the 1/f noise is currently unknown although efforts to determine it have ruled out potential sources including the MCT material. We plan to mitigate the 1/f by spatially modulating the signal on the detector with an internal pupil-plane chopping mirror. However, when the device is operated at near full-well, the shot noise may approach or even dominate over the 1/f noise at low frequencies, suggesting the possibility of sampling the data at low frequency using only nodding. At 80% well with a 85 Hz frame rate, the 1/f noise is approximately 20% of the shot noise for a chopping (or nodding) frequency of 0.1 Hz, rising to 50% for a sampling frequency of 0.01 Hz.

The GeoSnap digitization electronics are integrated into the focal plane module, addressing each pixel individually. Signals are sent through a cold-to-warm cable through the cryostat wall, processed on two warm electronics boards, and then passed to a custom frame grabber board in a data acquisition PC which will also communicate with the telescope.

## 3. CRYOSTAT AND OPTICS

The MIRAC-5 cryostat is equipped with a PT405 Cryocooler from Cryomech. We expect to operate GeoSnap at temperatures between 35 K and 40 K while other parts of the MIRAC-5 interior are kept at 21 K. It takes approximately 24 hours for the interior of the cryostat to cool to these temperatures. The cold head is mounted directly to MIRAC-5. MIRAC-5's optical layout is designed to accommodate adaptive optics operations on the MMT and Magellan telescopes, specifically for multi-filter high contrast imaging, and perhaps future upgrades such as long-slit grism spectroscopy and coronagraphy.

An off-axis portion of an ellipsoid is used to reimage the telescope's focal plane onto the detector and the telescope entrance pupil onto the cold stop. The system features an adjustable front focal distance (FFD) via a "trombone" slide. MIRAC-4 supported an adjustable back focal distance (BFD) as well, but the larger and stiffer cables of GeoSnap disrupt this feature. As such the BFD is permanently positioned in MIRAC-4's "high magnification position." The 1024 pixel format of the detector and 18-micron pixel pitch results in a 19-arcsec field of view which is sufficient for our high-contrast imaging requirements. The system has a pixel scale of 0.019 arcseconds per pixel at MMT heavily oversampling the diffraction limit at 10 microns (0.317 arcseconds).

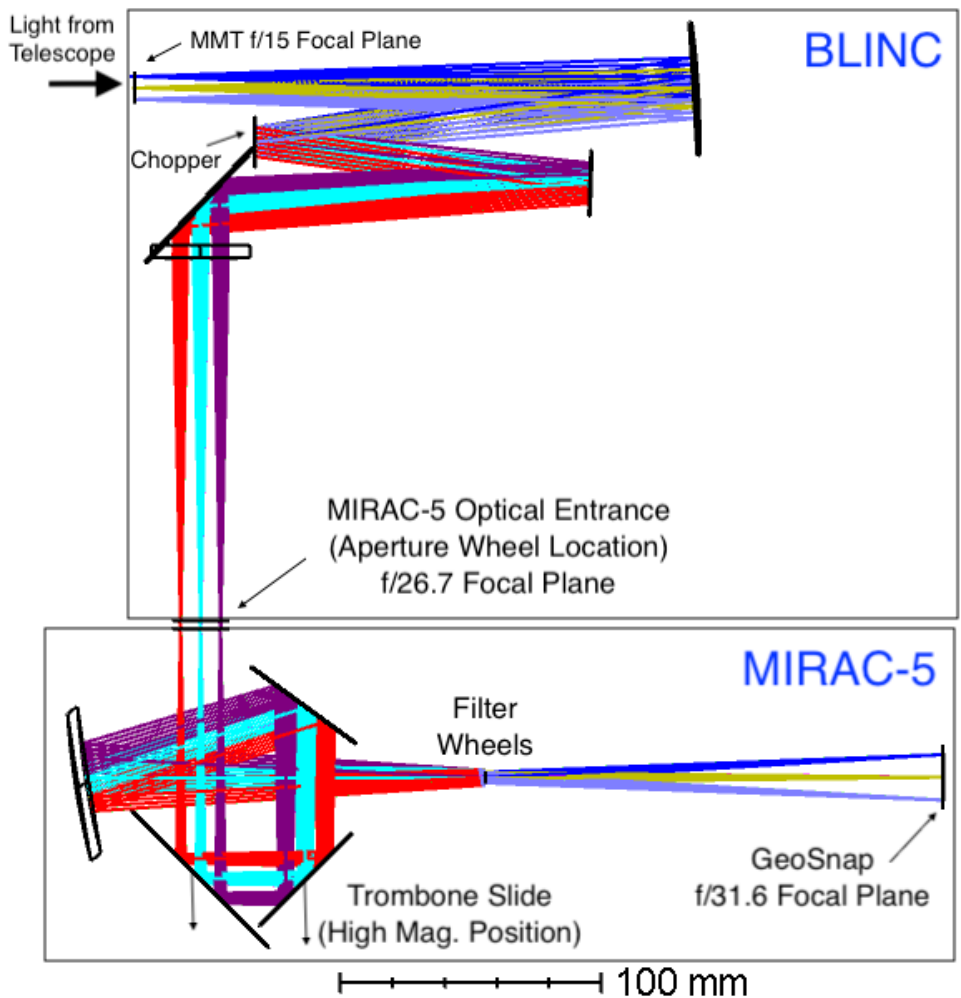

Figure 3: Optical layout of BLINC (top) and MIRAC-5 (bottom). A ray trace from the MMT is shown entering the system at the top left, passing through BLINC, entering MIRAC-5, and focusing on GeoSnap. Key locations in MIRAC-5 and f/#'s are marked.

MIRAC-5 is mated to BLINC, a separate system built for nulling interferometry[2]. BLINC is only used as re-imaging optics here, in order to get the correct f-number from the telescope into MIRAC. The optical layout of MIRAC-5 and BLINC is shown in Figure 3.

MIRAC-5 uses a pupil plane chopper inside BLINC which is driven by a rotational "voice coil" actuator. The chopper supports a beam angle of 1 degree which moves a source approximately half the FoV. At MMT, this will correspond to a 7.6 arcsec throw (24 $\lambda/D$ at 10 microns). The signal to the chopper is designed to cause acceleration, then drift, then deacceleration, followed by a hold level. This results in rapid transition (~1 msec compared to frame times of 10 msec) with minimum overshoot/ringing, minimum impacts on the hard stops, and low power consumption. The waveform can be adjusted by changing the voltage level for each of the four actions and the transition width. A damping switch can be enabled for damping at the end of a transition. An accelerometer system provides an automatic offset to correct for gravitational torque which varies with telescope orientation.

MIRAC-5 is equipped with an aperture wheel, cold stop wheel, and two filter wheels. Details for each wheel are listed below:

| Number | Name | Size |
|---|---|---|
| 1 (Home) | Large Square | 25.4x25.4 mm |
| 2 | Pinhole | 0.5 mm |
| 3 | Large Slit - Center | 0.71x25.4 mm |
| 4 | Small Slit - Offset | 0.58x25.4 mm |
| 5 | Large Slit - Offset | 0.71x25.4mm |

Table 1: Aperture wheel settings. The aperture wheel is located at the entrance of MIRAC-5.

| Number | Name | Size |
|---|---|---|
| 1 (Home) | Pinhole | 0.51 mm |
| 2 | Magellan | 8.99 mm |
| 3 | Dual Pupil - Nominal | 2 x 1.78 mm |
| 4 | MMT | 6.08 mm |
| 5 | Dual Pupil - Oversize | 2 x 1.96 mm |
| 6 | Blank | |

Table 2: Pupil wheel settings.

| Number | Name | Central $\lambda$ (microns) | FWHM (microns) | $\lambda_1$ (microns) | $\lambda_2$ (microns) | Effective Response |
|---|---|---|---|---|---|---|
| 1 (Home) | Open | | | | | |
| 2 | L' | 3.84 | 0.62 | 3.53 | 4.15 | 0.89 |
| 3 | M' | 4.66 | 0.24 | 4.55 | 4.80 | 0.82 |
| 4 | N0790 | 7.93 | 0.70 | 7.57 | 8.27 | 0.83 |
| 5 | W0870 | 8.74 | 1.23 | 8.13 | 9.36 | 0.87 |
| 6 | N0915 | 9.15 | 0.80 | 8.76 | 9.56 | 0.80 |
| 7 | N0980 | 9.82 | 0.92 | 9.35 | 10.27 | 0.83 |
| 8 | W1055 | 10.57 | 0.97 | 10.08 | 11.05 | 0.86 |
| 9 | N1185 | 11.89 | 1.14 | 11.32 | 12.46 | 0.81 |
| 10 | N1252 | 12.55 | 1.17 | 11.96 | 13.13 | 0.78 |
| 11 | N' | 11.34 | 2.27 | 10.22 | 12.49 | 0.84 |
| 12 | Ammonia* | 10.59 | 0.64 | 10.27 | 10.91 | 0.80 |

Table 3: Filter Wheel 1. The central wavelengths (Central $\lambda$), the full-width half-maximum (FWHM), the two 50% cutoff wavelengths ($\lambda_1$, $\lambda_2$), and the effective response are given. (*Ammonia filter is currently being produced; estimated specifications subject to change)

| Number | Name | Central $\lambda$ (microns) | FWHM (microns) | $\lambda_1$ (microns) | $\lambda_2$ (microns) | Effective Response |
|---|---|---|---|---|---|---|
| 1 (Home) | Open | | | | | |
| 2 | N-band | 10.85 | 5.71 | 8.00 | 13.71 | 0.88 |
| 3 | BaF$_2$ Blocker | Blocking at greater than 10 microns | | | | |
| 4 | EO 64355 | 10% throughput in observable wavelengths | | | | |
| 5 | H-band | 1.59 | 0.41 | 1.38 | 1.79 | 0.75 |
| 6 | K-band | 2.22 | 0.35 | 2.03 | 2.38 | 0.64 |
| 7 | M-band | 4.76 | 0.59 | 4.47 | 5.06 | 0.87 |
| 8 | Blank | | | | | |
| 9 | Blank | | | | | |
| 10 | Pupil Imaging Lens (New) | | | | | |
| 11 | Pupil Imaging Lens (Old) | | | | | |
| 12 | Blank | | | | | |

Table 4: Filter wheel 2. The central wavelengths (Central λ), the full-width half-maximum (FWHM), the two 50% cutoff wavelengths ($\lambda_1$, $\lambda_2$), and the effective response are given. The Pupil Imaging Lenses can be used to focus the pupil wheel on the detector for alignment purposes.

The wheels can be operated simultaneously, allowing for usage of filter combinations. The aperture (focal plane) wheel is located at the entrance to MIRAC-5 while the two filter wheels are located before the detector (Figure 3).

## 4. AMMONIA FILTER

The scientific opportunities for MIRAC-5 include observing forming protoplanets and targeting known gas giants and brown dwarf companions for mid-IR continuum detection. Furthermore, companions can be searched for evidence of $NH_3$ in their atmospheres, assuming all other physical parameters are self-consistent. Using HELIOS[11] and PETIT[12] atmospheric models originally assembled for the Bern Evolutionary Models[13], we studied the modeled spectra of several promising MIRAC-5 targets including HR 8799c, Kappa And b, and GJ 504b. Varying the temperature, surface gravity, metallicity, and cloud model (for PETIT) we determined the optimal fit for a filter targeting the ~10.5 micron ammonia feature. We focus on GJ 504b whose surface temperature of 550 K and log(g) value of 3 should be suitable for detecting ammonia[5]. We determine the shape of filter transmission needed to maximize the S/N in detecting that feature given only two photometric measurements: the existing N' filter and a new ammonia filter. The N' filter is used to calibrate throughput and the source solid angle.

We used the APOLLO[14] code to generate spectra with and without ammonia present[15]. This was used to identify two major absorption bands at 10.4 and 10.8 microns. Then, using the HELIOS and PETIT models, we searched 9.80 to 10.94 microns and employed two separate filter optimization routines. In general, the two methods agreed, returning consistent outputs for the most favorable filter location (Figure 4).

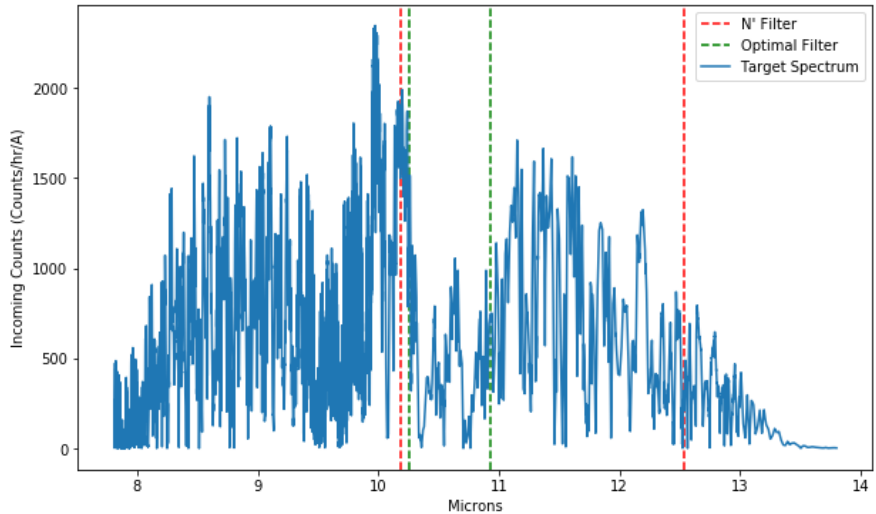

Figure 4: An example from the ammonia optimization routine for a 500 K, log(g) = 3.0, HELIOS target. The boundaries of the two filters (the N' filter and the optimized ammonia filter) are shown. A S/N of 3 can be obtained with 8 hours on this target in these two filters.

After surveying our parameter space, we determined the filter should span 10.27 to 10.91 microns. For a 500 K and log(g) = 3.0 GJ504b, the detection of the ammonia feature reaches a S/N of 3 in approximately 8 hours (total time in both filters). The primary driver of exposure time was the assumed temperature which we varied from 300 to 600 K. At 600 K, we found the required exposure time quickly began to rise, in a few cases requiring approximately 15 hours. Finally, for the purposes of ease of manufacturing (lower cost), we estimated the impact on the S/N for the EW if the edges of the filter are shifted by +-0.05 microns, our requested tolerance. We find the impact on the S/N ratio is only a few percent at that level of uncertainty.

## 5. FUTURE DEVELOPMENTS AND CLOSING REMARKS

MIRAC-5 will be commissioned on MMT with the MAPS AO[3] system near the end of 2022. Eventually, MIRAC-5 will also be used at Magellan with MagAO[4]. Already, MIRAC-5 has the potential for characterizing known gas giants and brown dwarf companions. However, higher contrast will be required to fully utilize the potential of MIRAC-5. Following the successful commissioning of MIRAC-5 at MMT, we will plan for a future upgrade with an Annular Groove Phase Mask (AGPM)[16] coronagraph in the intermediate focal plane with associated optimized pupil mask. The AGPM will be supported by the implementation of a Quadrant Analysis of Coronagraphic Images for Tip-tilt Sensing (QACITS)[17] control loop with MAPS. These upgrades should make possible the observation of several known targets including Eps Indi A b or 51 Eri b, RV discovered long-period planets around mature stars, and a wide range of forming protoplanets[18].

## Acknowledgements

This project was funded by the Heising-Simons Foundation through grant 2020-1699. We are also grateful to Oli Durney for his assistance with optical design and Arthur Adams for his assistance with ammonia identification. Finally, we acknowledge encouragement and support from Teledyne Imaging Sensors, in particular Vincent Douence, John Auyeung, and Jim Beletic.## REFERENCES

[1] Hoffmann et al., "MIRAC2: a mid-infrared array camera for astronomy," Proc. SPIE 3354, 647-658 (1998).
[2] Hinz et al., "BLINC: a testbed for nulling interferometry in the thermal infrared," Proc. SPIE 4006, 349-353 (2000).
[3] Anugu et al., "Design and development of a high-speed visible pyramid wavefront sensor for the MMT AO system," Proc. SPIE 11448, 114485J (2020).
[4] Morzinski et al., "MagAO: status and science," Proc. SPIE 9909, 990901 (2016).
[5] Skemer et al., "The LEECH Exoplanet Imaging Survey: Characterization of the Coldest Directly Imaged Exoplanet, GJ 504 b, and Evidence for Superstellar Metallicity," ApJ 817, 2 (2016).
[6] Leisenring et al. In Prep.
[7] Cabrera et al., "Development of 13-μm cutoff HgCdTe detector arrays for astronomy," J. Astron. Telesc. Instrum. 5, 036005 (2019).
[8] McMurtry et al., "Development of sensitive long-wave infrared detector arrays for passively cooled space missions," Optical Engineering 52, 9 (2013).
[9] Lord, Steven D., "A new software tool for computing Earth's atmospheric transmission of near- and far-infrared radiation," NASA Technical Memorandum, 103957 (1992).
[10] Ives et al., "AQUARIUS: the next generation mid-IR detector for ground-based astronomy, an update," Proc. SPIE 9154, 488-499 (2014).
[11] Malik et al., "HELIOS: An Open-source, GPU-accelerated Radiative Transfer Code for Self-consistent Exoplanetary Atmospheres," ApJ 153, 56 (2017).
[12] Mollière et al., "petitRADTRANS: a Python radiative transfer package for exoplanet characterization and retrieval," A&A 627, A67 (2019).